\documentclass[12pt]{article}
\begin{document}

\title{Heading towards chaos criterion for quantum field systems}

\author{V.I. Kuvshinov$^{1}$, A.V. Kuzmin$^{1}$ \\
       1 -- Institute of Physics, NASB.\\
       E-mail:kuvshino@dragon.bas-net.by, \\
              avkuzmin@dragon.bas-net.by}

\maketitle

\begin{abstract}
Chaos criterion for quantum field theory is proposed. Its accordance with classical
chaos criterion is demonstrated in the semi-classical limit of quantum mechanics.
\end{abstract}

Originally phenomenon of chaos was associated with problems of classical mechanics and
statistical physics. Attempt of substantiation of statistical mechanics initiated
intensive study of chaos and uncover its basic properties mainly in classical mechanics.
One of the main results in this direction is a creation of KAM theory and understanding
of the phase space structure of Hamiltonian systems \cite{Kolmogorov}. It was clarified
that the root of chaos is local instability of dynamical system \cite{Krylov}. Local
instability leads to mixing of trajectories in phase space and thus to non-regular
behavior of the system and chaos \cite{Zaslavski}. Significant property of chaos is its
prevalence in various natural phenomena. It explains the importance of study of chaos
and a large number of works in this field.
\par
Large progress is achieved in understanding of chaos in semi-classical regime of quantum
mechanics via analysis of the spectral properties of the system
\cite{Zaslavski},\cite{Robnic}. Semi-classical restrictions are important, because a
large number of energy levels in small energy interval is needed to provide a statistics
\cite{Robnic2}.
\par
Investigation of the stability of classical field solutions faces difficulties caused by
infinite number of degrees of freedom. That is why authors often restrict their
consideration by the investigation of some model field configurations \cite{Kawabe}.
\par
There are papers devoted chaos in quantum field theory \cite{Q2}. But there is no
generally recognized definition of chaos for pure quantum systems \cite{Bunakov}. This
fact restricts use of chaos theory in the field of elementary particle physics. At the
same time it is well known that the field equations  of all four types of fundamental
interactions have chaotic solutions \cite{Int4} and high energy physics reveals the
phenomenon of intermittency \cite{Intermittency}.

To clear the role of chaos in particle physics one has to learn how chaos affects on
observable quantities. To reach it, it is necessary to study chaos in quantum gauge
field theories, modern theories of particle interactions. But, as it was mentioned
above, it is still unclear what is chaos in quantum field theory and what phenomena is
it become apparent in? Thus additional chaos criterion formulated for purposes of
quantum field theory is needed. It has to satisfy two general requirements. Primarily,
it has to agree with well known classical chaos criteria \cite{Zaslavski},
\cite{Shuster}. Secondly, modern language of particle physics has to be used in its
formulation to provide its direct application to quantum field systems. From our point
of view, path integrals formalism has to be used.

Now we give some qualitative arguments which bring us to the formulation of the chaos
criterion for quantum field theory. The language of path integrals let us to achieve
mathematical description of both quantum mechanics and quantum field theory in the
framework of the same formalism. It lets us to monitor the way from classical chaos
through quantum mechanics to the chaos in quantum field theory and thus to substantiate
the agreement of proposed chaos criterion with classical ones. Quantum mechanics is
needed to play the role of the bridge. From statistical mechanics and ergodic theory it
is known that chaos in classical systems is a consequence of the property of mixing
\cite{Zaslavski}. Mixing means rapid (exponential) decreasing of correlation function
with time \cite{Zaslavski}. Thus rapid (exponential) decrease of correlation function is
the signature of chaos \cite{Shuster}. In other words, if correlation function
exponentially decreases than the corresponding motion is chaotic, if it oscillates or is
constant then the motion is regular \cite{Shuster}. We expand criterion of this type for
quantum field systems.  All stated bellow remains valid for quantum mechanics, since
mathematical description via path integrals is the same.
\par
For field systems the analogue of the classical correlation function is two-point
connected Green function
\begin{equation}\label{green}
G_{ik}(x,y)= -\frac{\delta^{2}W[\vec{J}]}{\delta J_{i}(x)
 \delta J_{k}(y)}\left.\right|_{\vec{J}=0}
\end{equation}
Here $W[\vec{J}]$ is generating functional of connected Green functions, $\vec{J}$ are
the sources of the fields, $x$, $y$ are 4-vectors of space-time coordinates.
\par
Thus we formulate chaos criterion for quantum mechanics and quantum field theory in the
following form:
\par
a) If two-point Green function (\ref{green}) exponentially goes to zero when the
distance between its arguments goes to infinity then the system is chaotic.
\par
b) If it oscillates or remains constant in this limit then we have regular behavior of
the quantum system.
\par

Before to check the agreement with classical chaos criteria we need to make a step
aside. All classical chaos criteria in some way are based on the condition of the local
instability of classical trajectories \cite{Zaslavski}. We need to represent these
conditions in the form convenient for our purposes. In order to do it we generalize Toda
criterion of local instability \cite{Toda}, \cite{Ukraine}.

Toda criterion of chaos for classical mechanical systems was at first formulated in
\cite{Toda}.
 It was reformulated for Hamiltonian systems with two degrees of freedom by
  Salasnich \cite{Salasnich}. Agreement between Toda criterion and
criterion of classical chaos based on KAM theory and conception of nonlinear resonance
was checked on the particular example in \cite{we3}.
\par
Consider classical Hamiltonian system with any finite number of degrees of freedom
\begin{equation}
H=\frac{1}{2}\vec{p}^{2}+V(\vec{q}),\quad
\vec{p}=(p_{1,\ldots,}p_{N})\;;\;\vec{q}=(q_{1,\ldots ,}q_{N})
\end{equation}
Behavior of the classical system is locally unstable if distance between two neighboring
trajectories grows exponentially with time in some region of the phase space.
\par
 Linearized Hamilton equations are
\begin{equation}\label{lineareq}
\frac{d}{dt}\left(
\begin{array}
[c]{l}%
\delta\vec{q}\\
\delta\vec{p}%
\end{array}
\right)  = G  \left(
\begin{array}
[c]{l}%
\delta\vec{q}\\
\delta\vec{p}%
\end{array}
\right)  , \quad G\equiv\left(
\begin{array}
[c]{cc}%
0 & I\\
-\Sigma & 0
\end{array}
\right)
\end{equation}
Here $I$ is the $N \times N$ identity matrix, $G$ is a stability matrix and matrix $\Sigma$ is%
\begin{equation}\label{Sigma}
\Sigma\equiv\left(  \frac{\partial^{2}V}{\partial q_{i}\partial q_{j}}\left|
_{\vec{q}_{0}}\right.  \right)
\end{equation}
Matrix $\Sigma$ and stability matrix $G$ are functions of the point $\vec{q}_{0}$ of
configuration space of the system. Solution of the equations (\ref{lineareq}) valid in
$\Omega$ has the following form
\begin{equation}\label{solution}
\left(
\begin{array}
[c]{l}%
\delta\vec{q}(t)\\
\delta\vec{p}(t)
\end{array}
\right)  =\sum_{i=1}^{2N}C_{i}\exp\left\{  \lambda_{i}t\right\} \left(
\begin{array}
[c]{l}%
\delta\vec{q}(0)\\
\delta\vec{p}(0)
\end{array}
\right)
\end{equation}
Here $\lambda_{i}=\lambda_{i}(\vec{q}_{0})$ are eigenvalues of the stability matrix $G$.
And $\{C_{i}\}$ is a full set of projectors. From (\ref{solution}) it is seen:
\par
a) If there is $i$ such as $Re\lambda _{i}\neq0$ then the distance between neighboring
trajectories grows exponentially with time and the motion is locally unstable.
\par
b) If for any $i=\overline{1,2N}\quad$ $Re\lambda_{i}=0$ \ then there is no local
instability and the motion is regular.
\par
It is easy to see that $G^{2}=diag(-\Sigma, -\Sigma)$.Therefore if
$(-\xi_{i})\;,\;i=\overline{1,N}$ are eigenvalues of the matrix $(-\Sigma)$ then
\begin{equation}
(-\xi_{i})=\lambda_{i}^{2}\;,\;i=\overline{1,N} \quad ,
 \quad \lambda_{i}^{2}=\lambda_{i+N}^{2}\;,\;i=\overline{1,N}%
\end{equation}
Thus without loss of generality we can imply that $Re\lambda_{i}\geq0$. Notice that
\begin{equation}\label{123}
\xi_{i}=-\lambda_{i}^{2}=(Im\lambda_{i})^{2}-
(Re%
\lambda_{i})^{2}- 2iIm\lambda_{i}Re\lambda _{i}
\end{equation}
Since matrix $\Sigma$ is real and symmetric its eigenvalues $\{\xi _{i}\} ,\quad
i=\overline{1,N}$ are real. Therefore (\ref{123}) leads to the condition
$Im\lambda_{i}Re\lambda_{i}=0, \quad
 \forall i=\overline{1,N}$
Thus any eigenvalue of the stability matrix $G$ is real or pure imaginary or equals
zero. Therefore the generalized Toda criterion for classical Hamiltonian systems with
any finite number of freedoms can be formulated as follows:
\par
a) If $\xi_{i}\geq0 \quad , \quad \forall i=\overline{1,N}$ \ then behavior of the
system is regular near the point $\vec{q}_{0}$.
\par
b) If $\exists i=\overline{1,N}:\xi_{i}<0$ then behavior of the system is locally
unstable near the point $\vec{q}_{0}$.
\par
If one of these conditions holds in some region of the configuration space then the
motion is stable or chaotic respectively in this region. These results in the case of
the system with two degrees of freedom coincide with ones obtained in \cite{Salasnich}.

To check the agreement between generalized Toda criterion and formulated quantum chaos
criterion in semi-classical limit we shall calculate two-point Green function in
semi-classical approximation of quantum mechanics. Generating functional is
\begin{equation}
Z[\vec{J}]= \int D\vec{q} \exp{\{i \int^{+\infty}_{- \infty} dt \left[
\frac{1}{2}\dot{\vec{q}}^{2} - V(\vec{q}) + \vec{J}\vec{q} \right] \}}
\end{equation}
Consider certain solution of classical equations of motion $\vec{q}_{0}(t)$. Introduce
new variable describing the deviations from the classical trajectory $
\delta\vec{q}(t)=\vec{q}-\vec{q}_{0}(t)$, then under semi-classical approximation
$$
Z[\vec{J}] = \exp{\{i S_{0}[\vec{J}]\}} \int D\delta\vec{q} exp\{i
\int^{+\infty}_{-\infty} dt \left[\frac{1}{2}\delta\dot{\vec{q}}^{2} \right.-
$$
\begin{equation}\label{GF}
-\left. \frac{1}{2}\delta\vec{q} \Sigma \delta\vec{q} + \vec{J}\delta\vec{q} \right] \}
\end{equation}
Here classical action
\begin{equation}
S_{0}[\vec{J}] = \int^{+\infty}_{-\infty} dt \left[\frac{1}{2} \dot{\vec{q}}_{0}^{2} -
V\left(\vec{q}_{0}(t)\right) + \vec{J}\vec{q}_{0} \right]
\end{equation}
and matrix $\Sigma$ is defined by (\ref{Sigma}). Here for simplicity we suppose that
$\Sigma$ does not depend on time and remains constant on the classical trajectory.
Matrix $\Sigma$ is real and symmetric, therefore diagonalizing orthogonal matrix does
exist. Thus after the orthogonal transformation we get Gaussian path integrals.
Calculation in this case gives following Green function
\begin{equation}\label{MainGreen2}
 G_{i}(t_{1},t_{2})=\frac{i}{2}Re\left( \frac{e^{-\lambda_{i}(t_{1} -
 t_{2})}}{\lambda_{i}} \right)
\end{equation}
From the expression (\ref{MainGreen2}) it is seen
\par
a) If classical motion is locally unstable (chaotic) then according Toda criterion there
is real eigenvalue $\lambda_{i}$ and therefore Green function (\ref{MainGreen2})
exponentially goes to zero for some $i$. Opposite is also true. If Green function
(\ref{MainGreen2}) exponentially goes to zero under the condition
$|t_{1}-t_{2}|\rightarrow \infty$ for some $i$, then there exists real eigenvalue of the
stability matrix and thus classical motion is locally unstable.
\par
b) If all eigenvalues of the stability matrix $G$ are pure imaginary, that corresponds
classically stable motion, then in the limit $|t_{1}-t_{2}|\rightarrow \infty$ Green
function (\ref{MainGreen2}) oscillates as a sine. Opposite is also true. If for any $i$
Green functions oscillate in the limit $|t_{1}-t_{2}|\rightarrow \infty$ then
$\{\lambda_{i}\}$ are pure imaginary for any $i$ and classical motion is stable and
regular.
\par
Thus we have demonstrated that proposed quantum chaos criterion coincides with Toda
criterion in the semi-classical limit (corresponding principle).
\par

We have generalized Toda criterion for the Hamiltonian systems with any finite number of
degrees of freedom. Basing on the formal analogy between statistical mechanics and
quantum field theory we proposed chaos criterion for quantum mechanical and quantum
field systems. Consideration of quantum mechanics is needed to provide a bridge from
classical chaos to chaos in quantum field theory. We have demonstrated that proposed
chaos criterion corresponds to generalized Toda criterion in semi-classical limit of
quantum mechanics in the case when Lyapunov exponents do not depend on time.

\end{document}